\documentstyle[12pt]{article}

\begin{document}
\setcounter{page}{1}
\pagestyle{plain} \vspace{1cm}
\begin{center}
\Large{\bf Dynamical-Screening and the Phantom-Like Effects in a
DGP-Inspired $F(R,\phi)$ Model}\\
\small \vspace{1cm}
{\bf Kourosh Nozari$^{*}$}\quad\ and \quad {\bf Faeze Kiani$^{\dag}$}\\
\vspace{0.5cm} {\it Department of Physics,
Faculty of Basic Sciences,\\
University of Mazandaran,\\
P. O. Box 47416-95447, Babolsar, IRAN}\\
{\it $^{*}$ knozari@umz.ac.ir}\\
{\it $^{\dag}$ fkiani@umz.ac.ir}
\end{center}
\vspace{1.5cm}
\begin{abstract}
Based on the Lue-Starkman conjecture on the dynamical screening of
the brane cosmological constant in the DGP scenario, we extend this
proposal to a general DGP-inspired $F(R,\phi)$ Model. We show that
modification of the induced gravity and its coupling to a
quintessence field localized on the brane, affects the screening of
the brane cosmological constant and also phantom-like behavior on
the brane. We extend our study to possible modification of the
induced gravity on the brane and for clarification some specific
examples are presented. As a result, phantom-like behavior can be
realized in this setup without violating the null energy condition
at least in some subspaces of the model parameter space. The key
result of our study is the fact that a DGP-inspired $F(R,\phi)$
scenario has the best fit with LCDM and recent observations than
other alternative theories.
\end{abstract}
\vspace{1.5cm}
\newpage

\section{Introduction}
Recent evidences from supernova searches data [1,2], cosmic
microwave background (CMB) results [3-5] and also Wilkinson
Microwave Anisotropy Probe (WMAP) data [6,7], show an positively
accelerating phase of the cosmic expansion today and this feature
shows that the simple picture of the universe consisting of the
pressureless fluid is not enough to describe the cosmological
dynamics. In this regard, the universe may contain some sort of the
additional negative-pressure dark energy. Analysis of the WMAP data
[8-10] shows that there is no indication for any significant
deviations from Gaussianity and adiabaticity of the CMB power
spectrum and therefore suggests that the universe is spatially flat
to within the limits of observational accuracy. Further, the
combined analysis of the WMAP data with the supernova Legacy survey
(SNLS) [8], constrains the equation of state $w_{de}$, corresponding
to almost ${74\%}$ contribution of dark energy in the currently
accelerating universe, to be very close to that of the cosmological
constant value. In this respect, a LCDM ( Cosmological constant plus
Cold Dark Matter) model has maximum agreement with the recent data.
Moreover, observations appear to favor a dark energy equation of
state, $w_{de}<-1$ [11]. Therefore, a viable cosmological model
should admit a dynamical equation of state that might have crossed
the value $w_{de}= -1$ in the recent epoch of cosmological evolution
[12]. In fact, to explain positively accelerated expansion of the
universe, there are two alternative approaches: incorporating an
additional cosmological component ( dark energy) in matter sector of
the general theory of relativity ( $G_{\mu\nu}=8\pi G
(T^{(M)}_{\mu\nu}+T^{(Dark)}_{\mu\nu})$\, where $T^{(M)}_{\mu\nu}$
and $T^{(Dark)}_{\mu\nu}$ are energy-momentum tensor of ordinary
matter and dark energy respectively), or modifying geometric sector
of the theory (dark geometry)($G_{\mu\nu}+G^{Dark}_{\mu\nu}=8\pi G
T^{(M)}_{\mu\nu}$) at the cosmological scales. Multi-component dark
energy with at least one non-canonical {\it phantom} field is a
possible candidate of the first alternative. This viewpoint has been
studied extensively in the literature ( see [13,14] and references
therein ). Another alternative to explain current accelerated
expansion of the universe is extension of the general relativity to
more general theories on cosmological scales. In this view point,
modified Einstein-Hilbert action via $f(R)$-gravity ( see [15] and
references therein) or braneworld gravity [16-18] are studied
extensively. In this framework the geometric part of the Einstein's
field equations are modified. For instance, DGP (
Dvali-Gabadadze-Porrati) braneworld scenario as an IR modification
of the general relativity explains accelerated expansion of the
universe in its self-accelerating branch via leakage of gravity to
extra dimension. In this model, equation of state parameter of dark
energy never crosses the $\omega(z)=-1$ line, and universe
eventually turns out to be de Sitter phase. Nevertheless, in this
setup if we use a single scalar field (ordinary or phantom) on the
brane, we can show that equation of state parameter of dark energy
can cross phantom divide line [19]. One important consequence in the
quintessence model of dark energy is the fact that a single
minimally coupled scalar field has not the capability to explain
crossing of the phantom divide line, $\omega_{\phi}=-1$ [20].
However, a single but non-minimally coupled scalar field is enough
to cross the phantom divide line by its equation of state parameter
[13,14]. Lorentz invariance violating vector fields in an
interactive basis are other possibility to realize cosmological line
crossing [21]. Lue and Starkman [22] based on the analysis firstly
reported by Sahni and Shtanov [23] have shown that one can realize
the phantom-like effect ( increasing of the effective dark energy
density with cosmic time) in the normal branch of the DGP
cosmological solution without introducing any phantom field. This
type of the analysis then has been extended by several authors [24].
The normal branch of the model which cannot explain the
self-acceleration, has the key property that brane is extrinsically
curved so that shortcuts through the bulk allow gravity to screen
the effects of the brane energy-momentum contents at Hubble
parameters of the order of the inverse of crossover distance [22].
Since in this case $H(t)$ is a decreasing function of the cosmic
time, the effective dark energy component is increasing with time
and therefore we observe a phantom-like behavior without introducing
any phantom matter. It is important to note that crossing of the
phantom divide line in this viewpoint is impossible without
introduction of a quintessence field on the brane [24]. This idea
has been studied further to incorporate curvature effects [25]. The
importance of this type of reasoning lies in the fact that we don't
need to introduce phantom fields that violate the null energy
condition and suffer from several theoretical problems.

Here we are going to study phantom-like effect in the normal branch
of a general DGP inspired $F(R,\phi)$ scenario. The DGP inspired
$F(R,\phi)$ scenarios have been studied in Refs. [26,27]. Our
motivation to study phantom-like behavior of this extension of the
DGP scenario is the fact that to have crossing of the phantom divide
line on the DGP brane we have to incorporate a quintessence field on
the brane [24]. On the other hand, it is reasonable to assume that
induced gravity on the brane can be modified. In fact, as has been
argued in Refs [28], generalized version of DGP scenario ( such as
modified induced gravity), can be ghost free and can give rise to
transient acceleration ( see also [23] and [29]). Here we are focus
on the normal branch of the scenario which is ghost-free. We show
that for the case with $F(R,\phi)=\frac{1}{2}(1-\xi\phi^{2})R$, the
effective dark energy density reduces by increasing the values of
the non-minimal coupling, $\xi$. We extend our study to the general
$f(R)$-gravity to explore the role played by the modification of the
induced gravity on the screening of the brane cosmological constant
and the phantom-like effect. We show that phantom-like behavior can
be realized in this setup without violating the null energy
condition at least in some subspaces of the model parameter space.
The key result of our study is the fact that a DGP-inspired
$F(R,\phi)$ scenario has the best fit with LCDM and recent
observations.

\section{Non-minimal DGP Cosmology}
\subsection{The Setup}
The action of the DGP scenario in the presence of a non-minimally
coupled scalar field on the brane can be written as follows [27]
\begin{equation}
S=\int d^{5}x\frac{m^{3}_{4}}{2}\sqrt{-g}{\cal R}+\Bigg[\int
d^{4}x\sqrt{-q}\bigg(\frac{m_{3}^{2}}{2}\alpha(\phi)
R[q]-\frac{1}{2} q^{\mu\nu} \nabla_{\mu}\phi\nabla_{\nu}\phi
-V(\phi) + m^{3}_{4}\overline{K}+ {\cal{L}}_{m}\bigg)\Bigg]_{y=0},
\end{equation}
where we have included a general non-minimal coupling $\alpha(\phi)$
\,in the brane part of the action( for an interesting discussion on
the possible schemes to incorporate NMC in the formulation of the
scalar-tensor gravity see [30,26], and also [31] for a braneworld
viewpoint).\, $m_{3}^{2}=(8\pi G)^{-1}$ and  $y$ is the coordinate
of the fifth dimension and we assume that brane is located at
$y=0$.\, $g_{AB}$ is five dimensional bulk metric with Ricci scalar
${\cal{R}}$, while $q_{\mu\nu}$ is induced metric on the brane with
induced Ricci scalar $R$.\, $\overline{K}$ is trace of the mean
extrinsic curvature of the brane defined as
\begin{equation}
\overline{K}_{\mu\nu}=\frac{1}{2}\,\,\lim_{\epsilon\rightarrow
0}\bigg(\Big[K_{\mu\nu}\Big]_{y=-\epsilon}+
\Big[K_{\mu\nu}\Big]_{y=+\epsilon}\bigg),
\end{equation}
and corresponding term in the action is York-Gibbons-Hawking term
[33] (see also [34]). The ordinary matter part of the action is
shown by the Lagrangian ${\cal{L}}_{m}\equiv
{\cal{L}}_{m}(q_{\mu\nu},\psi)+\frac{\Lambda}{8\pi G}$ where $\psi$
is matter field and corresponding energy-momentum tensor is
\begin{equation}
T_{\mu\nu}=-2\frac{\delta{\cal{L}}_{m}}{\delta
q^{\mu\nu}}+q_{\mu\nu}{\cal{L}}_{m},
\end{equation}
and \,$\Lambda$\,  is the brane cosmological constant. Note that we
assume that in addition to brane cosmological constant, there is
some quintessence scalar field localized on the brane to have a more
general framework and in order to realize phantom divide line
crossing. The pure scalar field Lagrangian,\,
${\cal{L}}_{\phi}=-\frac{1}{2} q^{\mu\nu}
\nabla_{\mu}\phi\nabla_{\nu}\phi -V(\phi)$,\,\,  yields the
following energy-momentum tensor
\begin{equation}
 \tau_{\mu\nu}=\nabla_\mu\phi\nabla_\nu\phi-\frac{1}{2}q_{\mu\nu}(\nabla\phi)^2
-q_{\mu\nu}V(\phi).
\end{equation}
The Bulk-brane Einstein's equations calculated from action (1) are
given by
$$m^{3}_{4}\left({\cal R}_{AB}-\frac{1}{2}g_{AB}{\cal
R}\right)+$$
\begin{equation}
m^{2}_{3}{\delta_{A}}^{\mu}{\delta_{B}}^{\nu}\bigg[\alpha(\phi)\left(R_{\mu\nu}-
\frac{1}{2}q_{\mu\nu}R\right)-\nabla_{\mu}\nabla_{\nu}\alpha(\phi)+
q_{\mu\nu}\Box^{(4)}\alpha(\phi)\bigg]\delta(y)
={\delta_{A}}^{\mu}{\delta_{B}}^{\nu}\Upsilon_{\mu\nu}\delta(y),
\end{equation}
where $\Box^{(4)}$ is 4-dimensional (brane) d'Alembertian and
$\Upsilon_{\mu\nu}=T_{\mu\nu}+\tau_{\mu\nu}$\,. This relation can be
rewritten as follows
\begin{equation}
m^{3}_{4}\left({\cal R}_{AB}-\frac{1}{2}g_{AB}{\cal R}\right)+
m^{2}_{3}\alpha(\phi){\delta_{A}}^{\mu}{\delta_{B}}^{\nu}\left(R_{\mu\nu}-
\frac{1}{2}q_{\mu\nu}R\right)\delta(y)=
{\delta_{A}}^{\mu}{\delta_{B}}^{\nu}{\cal{T}}_{\mu\nu}\delta(y)
\end{equation}
where ${\cal{T}}_{\mu\nu}$ is the total energy-momentum on the brane
defined as follows
\begin{equation}
{\cal{T}}_{\mu\nu}=m^{2}_{3}\nabla_{\mu}\nabla_{\nu}\alpha(\phi)-m^{2}_{3}
q_{\mu\nu}\Box^{(4)}\alpha(\phi)+\Upsilon_{\mu\nu},
\end{equation}
From (6) we find
\begin{equation}
G_{AB}={\cal R}_{AB}-\frac{1}{2}g_{AB}{\cal R}=0
\end{equation}
and
\begin{equation}
G_{\mu\nu}=\left(R_{\mu\nu}-
\frac{1}{2}q_{\mu\nu}R\right)=\frac{{\cal
T}_{\mu\nu}}{m^{2}_{3}\alpha(\phi)}
\end{equation}
for bulk and brane respectively. The corresponding junction
conditions relating the extrinsic curvature to the energy-momentum
tensor of the brane, have the following form
\begin{equation}
\lim_{\epsilon\rightarrow+0}\Big[K_{\mu\nu}\Big]^{y=+\epsilon}_{y=-\epsilon}
=\frac{1}{m_{4}^{3}}\bigg[{\cal{T}}_{\mu\nu}-\frac{1}{3}q_{\mu\nu}q^{\alpha\beta}
{\cal {T}}_{\alpha\beta}\bigg]_{y=0}
-\frac{m^{2}_{3}\alpha(\phi)}{m^{3}_{4}}\bigg[R_{\mu\nu}-
\frac{1}{6}q_{\mu\nu}q^{\alpha\beta}R_{\alpha\beta}\bigg]_{y=0}.
\end{equation}
Now we study cosmological dynamics in this setup. Since DGP scenario
accounts for embedding of the FRW cosmology at any distance scale
[33,34], we start with following line-element
\begin{equation}
ds^{2}=q_{\mu\nu}dx^{\mu}dx^{\nu}+b^{2}(y,t)dy^{2}=-n^{2}(y,t)dt^{2}+
a^{2}(y,t)\gamma_{ij}dx^{i}dx^{j}+b^{2}(y,t)dy^{2}.
\end{equation}
In this relation $\gamma_{ij}$ is a maximally symmetric
3-dimensional metric defined as
\begin{equation}
\gamma_{ij}=\delta_{ij}+k\frac{x_{i}x_{j}}{1-kr^{2}}
\end{equation}
where $k=-1,0,1$ parameterizes the spatial curvature and
$r^2=x_{i}x^{i}$. By computing components of Einstein's tensor and
using junction condition given in equation (10), we arrive at the
following Friedmann equation in this non-minimal DGP setup [27]
\begin{equation}
H^{2}+\frac{k}{a^2}=\frac{1}{3m_{3}^{2}\alpha(\phi)}
\bigg(\rho_{m}+\rho_{\phi}+\rho_{\Lambda}+\rho_{0}\bigg[1+\varepsilon
\sqrt{1+\frac{2}{\rho_{0}}
\Big[\rho_{m}+\rho_{\phi}+\rho_{\Lambda}-m_{3}^{2}\alpha(\phi)
\frac{{{\cal{E}}_{0}}}{a^{4}}\Big]}\,\,\bigg]\,\,\bigg).
\end{equation}
where $\rho_{0}=\frac{6m_{4}^{6}}{m_{3}^{2}\alpha(\phi)}$, \,
$\rho_{\Lambda}\equiv \frac{\Lambda}{8\pi
G}=m_{3}^{2}\alpha(\phi)\Lambda$ \, and $\rho_{m}$ is density of
ordinary matter on the brane.\, Also,  $\varepsilon=\pm1$ \, shows
the possibility of existence of two different branches of
DGP-inspired FRW equation corresponding to two different embedding
of the brane in the bulk. Neglecting the dark radiation term
$\frac{{{\cal{E}}_{0}}}{a^{4}}$ ( where ${\cal{E}}_{0}$ is an
integration constant) which decays very fast at late-times, we
rewrite equation (13) as follows
\begin{equation}
H^{2}=\frac{8\pi
G}{3}\Big(\rho_{m}+\rho_{\phi}\Big)+\frac{\Lambda}{3}
+\frac{1}{2r_{0}^{2}} + \varepsilon\sqrt{\frac{1}{4r_{0}^{4}}+
\frac{1}{r_{0}^{2}}\Big[\frac{8\pi
G}{3}\Big(\rho_{m}+\rho_{\phi}\Big)+\frac{\Lambda}{3}\Big]},
\end{equation}
where $r_{0}$ is a crossover distance defined as $r_{0}=\ell_{DGP}
\alpha(\phi)$\, and $G\equiv G_{eff}=1/8\pi m_{3}^{2}\alpha(\phi)$.
Now, the Friedmann equation (14) can be rewritten as follows
\begin{equation}
H^{2}=\frac{8\pi
G}{3}\Big(\rho_{m}+\rho_{\phi}\Big)+\frac{\Lambda}{3} +
\varepsilon\frac{H}{r_{0}}.
\end{equation}
We use this equation in our forthcoming arguments.\\

\subsection{Lue-Starkman Screening of the Brane Cosmological
Constant} As we have pointed out in the introduction, Lue and
Starkman have shown that one can realize phantom-like effect, that
is, increasing of the effective dark energy density with cosmic
time, in the normal branch of the DGP cosmological solution without
introducing any phantom field. The normal branch of the model which
cannot explain the self-acceleration, has the key property that
brane is extrinsically curved so that shortcuts through the bulk
allow gravity to screen the effects of the brane energy-momentum
contents at Hubble parameters $H\sim r_{0}^{-1}$\, where $r_{0}$ is
the crossover distance [22]. Since in this case $H(t)$ is a
decreasing function of the cosmic time, the effective dark energy
component is increasing with time and therefore we observe a
phantom-like behavior without introducing any phantom matter that
violate null energy condition and suffers from several theoretical
problems. In the first step, in this section we study the
phantom-like effect in the normal branch of a DGP inspired
\emph{non-minimal} scenario. In other words, here we suppose that
there is a quintessence field non-minimally coupled to the induced
gravity on the DGP brane. We emphasize that we have included a
canonical (quintessence) scalar field to incorporate possible
coupling of the gravity and scalar degrees of freedom on the brane.
This provides a wider parameter space with capability to handle the
problem more complete. In fact, inclusion of this field brings the
theory to realize crossing of the phantom divide line [24]. As has
been shown by Chimento {\it et al.}\,, the normal branch of the DGP
scenario has the capability to describe phantom-like effect but it
cannot realize crossing of the phantom divide line without
introducing a quintessence scalar field on the brane. With this
motivation, here we have considered the existence a canonical scalar
field on the brane that couples non-minimally with induced gravity.
In the next section we incorporate possible modification of the
induced gravity on the brane too.

Considering the normal branch of the equation (15) with
$\varepsilon=-1$, we have
\begin{equation}
H^{2}\approx\frac{8\pi
G}{3}\Big(\rho_{m}+\rho_{\phi}\Big)+\frac{\Lambda}{3}
-\frac{H}{r_{0}}.
\end{equation}
Comparing this equation with the following Friedmann equation
\begin{equation}
H^{2}=\frac{8\pi G}{3}\Big(\rho_{m}+\rho_{\phi}\Big)+\frac{8\pi
G}{3}\rho^{(eff)}_{DE},
\end{equation}
we find\footnote{Note that this comparison is not perfect since $G$
in equation (16) is an effective quantity defined as $G\equiv
G_{eff}=1/8\pi m_{3}^{2}\alpha(\phi)$. However, since screening of
the brane cosmological constant can be attributed just to the last
two terms of the right hand side of equation (16), this comparison
is actually possible.}
\begin{equation}
\frac{8\pi G}{3}\rho_{DE}^{(eff)}=\frac{\Lambda}{3}
-\frac{H}{r_{0}}.
\end{equation}
Existence of a quintessence field nonminimally coupled to the
induced gravity on the brane leads to a redefinition of the
crossover scale as $r_{0}=\ell_{DGP} \alpha(\phi)$. Using definition
of $r_{0}$, equation (18) can be rewritten as follows
\begin{equation}
\frac{8\pi G}{3}\rho_{DE}^{(eff)}=\frac{\Lambda}{3}
-\frac{\alpha^{-1}H}{\ell_{DGP}}.
\end{equation}
Now we assume a conformal coupling of the scalar field and induced
gravity as follows
\begin{equation}
\alpha(\phi)=\frac{1}{2}(1-\xi\phi^{2}).
\end{equation}
The values of the $\xi$ is constraint by the observations from
different viewpoints ( see for instance [35,36]). The division by
$1-\xi \phi^{2}$ in our field equations unavoidably introduces the
two critical values of the scalar field $\pm
\phi_{c}=\pm\frac{1}{\sqrt{\xi}}$,\, for $\xi>0$, which are barriers
that the scalar field cannot cross. Note that in these values, the
effective gravitational coupling, its gradient, and the total
stress-energy tensor diverge ( see [31] for more details).

Now, by adopting ansatz (20), equation (19) can be rewritten as
follows
\begin{equation}
\frac{8\pi
G}{3}\rho_{DE}^{(eff)}=\frac{\Lambda}{3}-\frac{2H}{\ell_{DGP}(1-\xi\phi^{2})}.
\end{equation}
Figure $1$ shows the variation of $\rho_{DE}^{(eff)}$ versus $\xi$
in a constant time slice. In plotting this figure we have used the
ansatz $a(t)=a_{0}t^{\nu}$, $\phi(t)=\phi_{0}t^{-\mu}$ with
$\nu=1.2$ ( an accelerating phase of expansion) and $\mu=0.9$ ( a
decreasing quintessence field). The range of $\xi$ are chosen from
[36] constraint by the recent observations. As this figure shows, by
increasing the values of the nonminimal coupling,
$\rho_{DE}^{(eff)}$ decreases in a fixed time slice.
\begin{figure}[htp]
\begin{center}\includegraphics{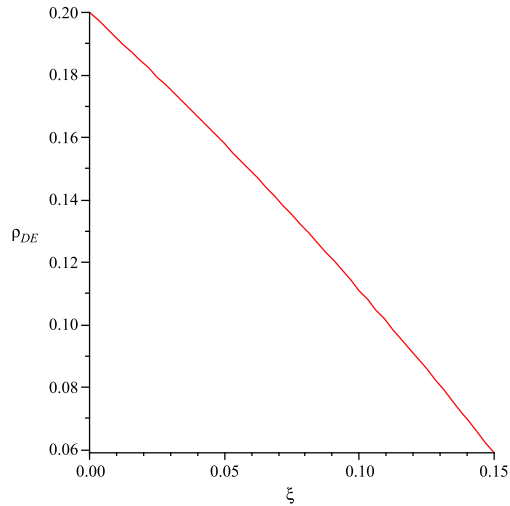} \vspace{7cm}
\end{center}
 \caption{\small {Variation of $\rho_{DE}^{(eff)}$ versus $\xi$ in a $t=constant$ slice.}}
\end{figure}
Figure $2$ shows the variation of the effective dark energy density
versus the cosmic time.
\begin{figure}[htp]
\begin{center}\includegraphics{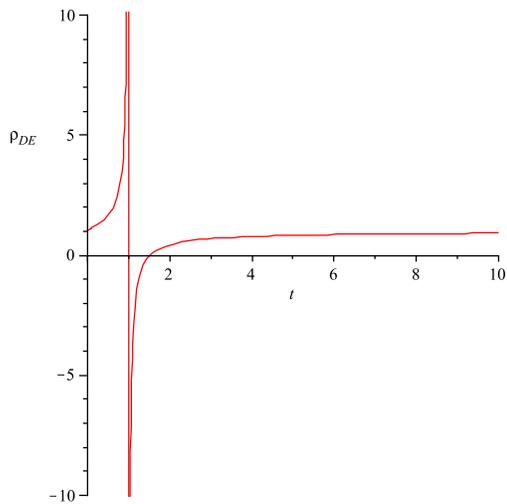} \vspace{5cm}
\end{center}
 \caption{\small { Variation of the effective dark energy density
versus the cosmic time. The effective dark energy density increases
with time and therefore shows a phantom-like behavior.}}
\end{figure}
As this figure shows, $\rho_{DE}^{eff}$ increases with cosmic time
and this is exactly the phantom-like behavior we are interested in.
Note that this phantom-like effects is realized without introducing
any phantom matter on the brane and only screening of the brane
cosmological constant causes such an intriguing effect. Although the
existence of a canonical scalar field non-minimally coupled to the
induced gravity on the brane has no considerable effect on the
phantom-like behavior but as figures $1$ and $3$ show, increasing
the values of the non-minimal coupling leads to the reduction of the
effective dark energy on a constant time slice. Figure $3$ shows the
variation of the effective dark energy versus the cosmic time and
non-minimal coupling. We note that while the introduction of a
phantom field requires the violation of the null energy condition,
here this energy condition is respected since we have not included
any phantom matter on the brane. Since the phantom-like dynamics
realized in this setup is gravitational ( the quintessence field
introduced here plays the role of standard matter on the brane), the
null energy condition cannot be violated in this case.
\begin{figure}[htp]
\begin{center}\includegraphics{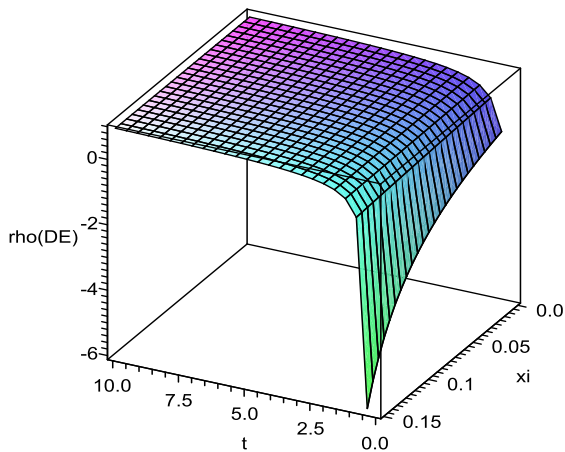} \vspace{6cm}
\end{center}
 \caption{\small {Variation of the effective dark energy versus
the cosmic time and the non-minimal coupling. For a constant $\xi$
the model realizes phantom-like effect versus cosmic time.}}
\end{figure}

\newpage
\section{DGP-inspired $F(R,\phi)$ Gravity}
\subsection{The Setup}
Now we extend our previous analysis to the more general case with
DGP-inspired $F(R,\phi)$ models. In other words, we incorporate
possible modification of the induced gravity on the brane. We assume
also a general coupling between a quintessence field localized on
the brane and modified induced gravity ( these types of theories
have been studied extensively and from various perspectives, see for
instance [26, 30, 37]). The action of this model is as follows
\begin{eqnarray}
S=\frac{m_{4}^{3}}{2}\int d^{5}(x)\sqrt{-g} \Re + \int
d^{4}(x)\sqrt{-q}\Big(\frac{m_{3}^{2}}{2}
F(R,\phi)-\frac{1}{2}q^{\mu\nu}\nabla_{\mu}\phi\nabla_{\nu}
\phi-V(\phi)+m_{4}^{3}\overline{K}+L_{m}\Big),
\end{eqnarray}
where the first term shows the usual Einstein-Hilbert action in 5D
bulk with 5D metric denoted by $g_{AB}$ and Ricci scalar denoted by
$\Re$. The second term on the right is a generalization of the
Einstein-Hilbert action induced on the brane. This is an extension
of the scalar-tensor theories in one side and a generalization of
$f(R)$-gravity on the other side. We call this model as DGP-inspired
$F(R,\phi)$ scenario. $y$ is the coordinate of the fifth dimension
and we suppose that brane is located at $y=0$ . $q_{\mu\nu}$ is
induced metric on the brane which is connected to $g_{AB}$ via
$q_{\mu\nu} = \delta_{\mu}\,^{A}\delta_{\nu}\,^{B}g_{AB}$. We denote
matter field Lagrangian by
$L_{m}=L_{m}(q_{\mu\nu},\psi)+\frac{\Lambda}{8\pi G}$ with
energy-momentum tensor defined as $T_{\mu\nu}=-2\frac{\delta
L_{m}}{\delta q^{\mu\nu}}+q_{\mu\nu}L_{m}$. The pure scalar field
lagrangian is $L_{\phi}=-\frac{1}{2}q^{\mu\nu} \nabla
_{\mu}\phi\nabla_{\nu}\phi-V(\phi)$ which gives the following
energy-momentum tensor
\begin{eqnarray}\tau_{\mu\nu}=\nabla_\mu  {\phi}  \nabla_{\nu} \phi -
\frac{1}{2} q_{\mu\nu}(\nabla \phi)^{2}-q_{\mu\nu}V(\phi).
\end{eqnarray}
The field equations resulting from this action are given as follows
\begin{equation}
\frac{m_{4}^{3}}{F'(R,\phi)}\bigg(\Re_{AB}-\frac{1}{2} g_{AB} \Re
\bigg)+m_{3}^{2}
 \delta_{A}\,^{\mu}\delta_{B}\,^{\nu} \Big(R_{\mu\nu}-
\frac{1}{2}q_{\mu\nu}R\Big)\delta(y)=
\delta_{A}\,^{\mu}\delta_{B}\,^{\nu}(\hat{T}_{\mu\nu}+\hat{\tau}_{\mu\nu}+
T_{\mu\nu}^{(curv)})\delta(y).
\end{equation}
In this relation
$\hat{T}_{\mu\nu}\equiv\frac{T_{\mu\nu}}{F'(R,\phi)}$ where
$T_{\mu\nu}$ is the energy-momentum tensor in matter frame and
$\hat{\tau}_{\mu\nu}\equiv\frac{\tau_{\mu\nu}}{F'(R,\phi)}$. A prime
denotes differentiation with respect to $R$. Also,
$T_{\mu\nu}^{(curv)}$ is defined as follows
\begin{equation}
T_{\mu\nu}^{(curv)}=\frac{m_{3}^{2}}{F'(R,\phi)}\bigg[
\frac{1}{2}q_{\mu\nu}\bigg(F(R,\phi)-RF'(R,\phi)\bigg
)+\bigg(F'(R,\phi)\bigg)^{;\alpha\beta}
\bigg(q_{\mu\alpha}q_{\nu\beta}-q_{\mu\nu}q_{\alpha\beta}\bigg)\bigg].
\end{equation}
In the bulk, $T_{AB}=0$  and therefore
\begin{eqnarray}G_{AB}=\Re_{AB}-\frac{1}{2}g_{AB}\Re=0
\end{eqnarray}
and on the brane we have
\begin{eqnarray}G_{\mu\nu}=R_{\mu\nu}-\frac{1}{2}q_{\mu\nu}R=
\frac{{\cal{T}}_{\mu\nu}}{m_{3}^{2}},
\end{eqnarray}
where $ {\cal{T}}_{\mu\nu}= \hat{T}_{\mu\nu}+\hat{\tau}_{\mu\nu}+
T_{\mu\nu}^{(curv)}$. The corresponding junction conditions relating
quantities on the brane are as follows
$$\lim_{\epsilon\longrightarrow+0}[K_{\mu\nu}]_{y=-\epsilon}^{y=+\epsilon}=
\frac{F'(R,\phi)}{m_{4}^{3}}\bigg[{\cal{T}}_{\mu\nu}-\frac{1}{3}q_{\mu\nu}q^{\alpha\beta}
{\cal{T}}_{\alpha\beta}\bigg]_{y=0}-$$
\begin{eqnarray}\frac{m_{3}^{2}}{m_{4}^{3}}F'(R,\phi)
\bigg[R_{\mu\nu}-\frac{1}{6}q_{\mu\nu}q^{\alpha\beta}R_{\alpha\beta}\bigg]_{y=0}\end{eqnarray}
A detailed study of weak field limit of this scenario within a
harmonic gauge on the longitudinal coordinates and using Green's
method to find gravitational potential, leads us to a modified
(effective) cross-over distance in this set-up as follows ( see [27]
for details of a similar argument)
\begin{eqnarray}\ell_{F}=\frac{m_{3}^{2}}{2m_{4}^{3}}
\bigg(\frac{dF}{dR}\bigg)=\bigg(\frac{dF}{dR}\bigg)\ell_{DGP},
\end{eqnarray}
where as usual $\ell_{DGP}=\frac{m_{3}^{2}}{2m_{4}^{3}}.$

\subsection{Cosmological Implications of the Model}

As we have explained in the previous section, embedding of FRW
cosmology in DGP setup is possible in the sense that this model
accounts for cosmological equations of motion at any distance scale
on the brane with any function of the Ricci scalar. To study
cosmology of a DGP-inspired $F(R,\phi)$ scenario, we consider the
line element as defined in equation (11). Also, we assume that the
scalar field $\phi$ depends only on the cosmic time on the brane.
Choosing a Gaussian normal coordinate system so that $b^{2}(y,t)=1$,
non-vanishing components of the Einstein's tensor in the bulk plus
junction conditions on the brane defined as
$$\lim_{\epsilon\longrightarrow+0}[\partial_{y}n]_{y=-\epsilon}^{y=+\epsilon}(t)=
\frac{2nm_{3}^{2}}{m_{4}^{3}}\bigg[
\Big(\frac{dF}{dR}\Big)\Big(\frac{\ddot{a}}{n^{2}a}-\frac{\dot{a}^{2}}{2n^{2}a^{2}}
-\frac{\dot{n}\dot{a}}{n^{3}a}-\frac{k}{2a^{2}}\Big)\bigg]_{y=0}$$
\begin{eqnarray}
+\frac
{n}{3m_{4}^{3}}\bigg[\Big(\frac{dF}{dR}\Big)\Big(2\rho^{(tot)}+3p^{(tot)}\Big)
\bigg]_{y=0},\end{eqnarray}
\begin{eqnarray}\lim_{\epsilon\longrightarrow+0}[\partial_{y}a]_{y=-\epsilon}^{y=+\epsilon}(t)=
\frac{m_{3}^{2}}{m_{4}^{3}}\bigg[\Big(\frac{dF}{dR}\Big)
\Big(\frac{\dot{a}^{2}}{n^{2}a}+\frac{k}{a}\Big)\bigg]_{y=0}-
\bigg[\bigg(\frac{dF}{dR}\bigg)\frac{\rho^{(tot)}a}{3m_{4}^{3}}\bigg]_{y=0}
\end{eqnarray}
yield the following generalization of the Friedmann equation for
cosmological dynamics on the brane ( see [26,27] for machinery of
calculations for a simple case)
\begin{eqnarray}
H^2+\frac{k}{a^2}=\frac{1}{3m_{3}^2F'(R,\phi)}\Bigg(\rho^{tot}
+\rho_{0}\bigg[1+\epsilon\sqrt{1+\frac{2}{\rho_{0}}\bigg[\rho^{tot}-
\frac{m_{3}^{2}F'(R,\phi)\varepsilon_{0}}{a^4}\bigg]}\bigg]\Bigg)
\end{eqnarray}
where $\epsilon=\pm1$ shows two different embedding of the brane, \,
$\rho_{0}=\frac{6m_{4}^{6}}{m_{3}^{2}F'(R,\phi)}$\, and
\,$\varepsilon_{0}=3\Big(\frac{\dot{a}^{2}}{n^{2}}-a'^{2}+k\Big)a^{2}$\,
is a constant with respect to $y$ ( with $a'\equiv\frac{da}{dy}$) (
see [34,27,38] for more detailed discussion on the constancy of this
quantity). Total energy density and pressure are defined as
~$\rho^{(tot)}=\hat{\rho}+\rho_{\phi}+\rho^{curv}+\rho_{\Lambda}$\,
and\, ~$p^{(tot)}=\hat{p}+p_{\phi}+p^{curv}+p_{\Lambda}$~
respectively. The ordinary matter on the brane has a perfect fluid
form with energy density $\hat{\rho}$ and pressure $\hat{p}$, while
the energy density and pressure corresponding to non-minimally
coupled quintessence scalar field and also those related to
curvature are given as follows
\begin{eqnarray}\rho_{\varphi}=\bigg[\frac{1}{2}\dot{\phi}^{2}+n^{2}V(\phi)-
6\frac{dF}{d\phi}H\dot{\phi}\bigg]_{y=0},\end{eqnarray}
\begin{eqnarray}p_{\phi}=\bigg[\frac{1}{2n^{2}}\dot{\phi}^{2}-V(\phi)+\frac{2}{n^{2}}
\frac{dF}{d\phi}(\ddot{\phi}-\frac{\dot{n}}{n}\dot{\phi})+
4\frac{dF}{d\phi}\frac{H}{n^{2}}\dot{\phi}+\frac{2}{n^{2}}
\frac{d^{2}F}{d\phi^{2}}\dot{\phi}^{2}\bigg]_{y=0}.\end{eqnarray}
also
\begin{eqnarray}\rho^{(curv)}=\frac{m_{3}^{2}}{F'(R,\phi)}\bigg(\frac{1}{2}
\bigg[F(R,\phi)-R F'(R,\phi)\bigg] -3\dot{R}H F''(R,\phi) \bigg),
\end{eqnarray}
\begin{eqnarray}p^{(curv)}=\frac{m_{3}^{2}}{F'(R,\phi)}\bigg({2\dot{R}H
F''(R,\phi)+\ddot{R}F''(R,\phi)
+\dot{R}^{2}F'''(R,\phi)-\frac{1}{2}\Big[ F(R,\phi)-R
F'(R,\phi)\Big]}\bigg).\end{eqnarray} where
$H=\frac{\dot{a}(0,t)}{a(0,t)}$~ is the Hubble parameter on the
brane. Ricci scalar on the brane is given by
$$R=3\frac{k}{a^2}+\frac{1}{n^{2}}\bigg[6\frac{\ddot{a}}{a}+
6\Big(\frac{\dot{a}}{a}\Big)^{2}-6\frac{\dot{a}}{a}\frac{\dot{n}}{n}\bigg].$$
Note that cosmological dynamics on the brane is given by setting
$n(0, t) = 1$. With this gauge condition we recover the usual time
on the brane via transformation $t = \int^{t} n(0,\eta)\,d\eta$
where $\eta$ is conformal time. It is interesting to note that the
equation of state parameter of the scalar field defined as
\begin{tiny}
\begin{eqnarray}\omega_{\phi}=\frac{\big(\frac{1}{2}+2\frac{d^2F}{d\phi^2}\big)\dot{\phi}^{2}-
V(\phi)+2\frac{d F}{d\phi}\big(\ddot{\phi}+2H\dot{\phi}\big)+
\frac{m_{3}^{2}}{F'(R,\phi)}\big[\big(2\dot{R}H+\ddot{R}\big)F''(R,\phi)+\dot{R}^2F'''(R,\phi)
-\frac{1}{2}F(R,\phi)+\frac{1}{2}R
F'(R,\phi)\big]}{\frac{1}{2}\dot{\phi}^{2}+V(\phi)- 6\frac{d
F}{d\phi}H\dot{\phi}+\frac{m_{3}^{2}}{F'(R,\phi)}\big(\frac{1}{2}F(R,\phi)-
\frac{1}{2}R F'(R,\phi)-3\dot{R}H\frac{d^2F}{dR^2}\big)},
\end{eqnarray}
\end{tiny}
crosses the phantom-divide line $\omega=-1$ in the favor of recent
observations [26,30].

\subsection{The Phantom-Like Behavior}

Now in this DGP-inspired $F(R,\phi)$ model, the crossover scale
takes the following form
\begin{equation}
\ell_{f}=\frac{m_{3}^{2}F'(R,\phi)}{2m_{4}^{3}}={F'(R,\phi)l_{DGP}},
\end{equation}
also
$$\rho_{0}=\frac{3m_{3}^{2}F'(R,\phi)}{2\ell_{f}^{2}}.$$
Neglecting the dark radiation term in equation (32), we find
\begin{equation}
H^2=\frac{8\pi
G(\rho_{m}+\rho_{\phi})}{3}+\frac{\Lambda}{3}+\frac{1}{2\ell_{f}^{2}}+
\varepsilon\sqrt{\frac{1}{4\ell_{f}^{4}}+\frac{1}{\ell_{f}^{2}}
\Big[{\frac{8\pi
G(\rho_{m}+\rho_{\phi})}{3}+\frac{\Lambda}{3}}\Big]}.
\end{equation}
where $G\equiv G_{eff}=\Big(8\pi m_{3}^{2}F'(R,\phi)\Big)^{-1}$ and
a prime denotes differentiation with respect to $R$. By adopting the
negative sign we find
\begin{equation}
H^2=\frac{8\pi
G(\rho_{m}+\rho_{\phi})}{3}+\frac{\Lambda}{3}-\frac{\Big[F'(R,\phi)\Big]^{-1}H}{l_{DGP}}.
\end{equation}
We can compare this equation with equation (17) to conclude that the
screening effect on the cosmological constant is modified as follows
\begin{equation}
\frac{8\pi
G}{3}\rho^{eff}_{DE}=\frac{\Lambda}{3}-\frac{\Big[F'(R,\phi)\Big]^{-1}H}{l_{DGP}}
\end{equation}
As an important especial case, for $F(R,\phi)=f(R)$ we find the
screening effect in a general $f(R)$-gravity
\begin{equation}
\frac{8\pi
G}{3}\rho_{DE}^{eff}=\lambda-\frac{\Big[f'(R)\Big]^{-1}H}{\ell_{DGP}}.
\end{equation}
Now as an enlightening example, we set for instance
$$F(R,\phi)=\frac{1}{2}(1-\xi\phi^{2})[R-(1-n)\zeta^{2}(R/\zeta^{2})^{n}]\,,
$$
where $\zeta$ is a suitably chosen parameter ( see for instance [15]
and [26]). With this choice, one recovers the general relativity if
$n=1$. For $n\neq1$, we obtain from equation (40)
\begin{equation}
\frac{8\pi
G}{3}\rho_{DE}^{eff}=\frac{\Lambda}{3}-\frac{2H}{\ell_{DGP}(1-\xi\phi^{2})(1-n(1-n)\zeta^{2(1-n)}R^{n-1})}
\end{equation}
For spatially flat FRW geometry the Riici scalar is given by
\begin{equation}
R=6\frac{\ddot{a}}{a}+6(\frac{\dot{a}}{a})^{2}.
\end{equation}
To have an intuition of phantom-like behavior in this case, we adopt
a suitable ansatz so that $a(t)= a_{0}t^{\nu}$ and $\phi(t)=\phi_{0}
t^{-\mu}$. We set $\nu=1.2$ and $\mu=0.9$ that are reliable from
physical grounds. Figure $4$ shows the variation of
$\rho_{eff}^{DE}$ versus $n$ in this DGP-inspired $F(R,\phi)$ model.
As we see, phantom-like behavior can be realized for $n\geq 0.73$
and $n\leq -0.60$. In other words, for $-0.6\leq n \leq 0.73$ the
effective dark energy in this DGP-inspired $F(R,\phi)$ model has no
phantom-like behavior.

\begin{figure}[htp]
\begin{center}\includegraphics{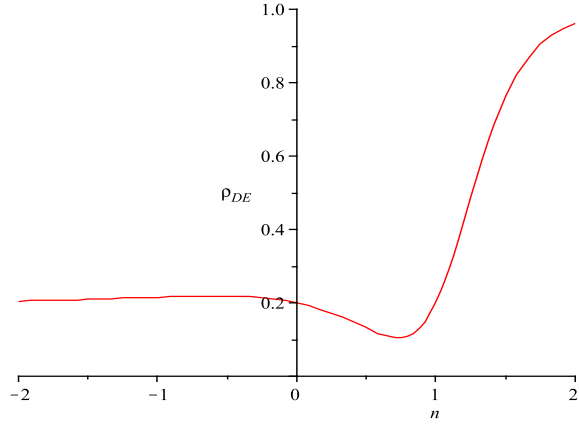} \vspace{4cm}
\end{center}
 \caption{\small {Variation of the effective dark energy versus
$n$ in a DGP-inspired $F(R,\phi)$ model with
$F(R,\phi)=\frac{1}{2}(1-\xi\phi^{2})[R-(1-n)\zeta^{2}(R/\zeta^{2})^{n}]
$. Phantom-like behavior can be realized for $n\geq 0.73$ and $n\leq
-0.60$.}}
\end{figure}
Figure $5$ shows the variation of the effective dark energy versus
$n$ and the non-minimal coupling. By increasing the values of $\xi$,
the effective dark energy density reduces but for a fixed value of
$\xi$, there is phantom-like effect for appropriate values of $n$.
\begin{figure}[htp]
\begin{center}\includegraphics{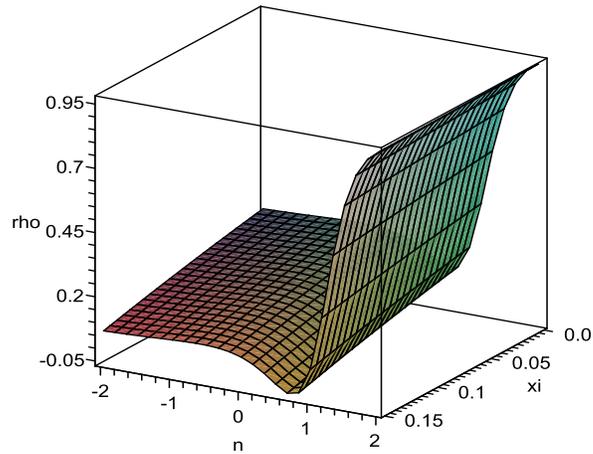} \vspace{5cm}
\end{center}
 \caption{\small {Variation of the effective dark energy versus
$n$ and the non-minimal coupling.}}
\end{figure}
Also, figure $6$ shows the variation of the effective dark energy
versus $n$ and the cosmic time. The phantom-like effect ( increasing
the values of the effective dark energy) can be realized for
suitable range of $n$.
\begin{figure}[htp]
\begin{center}\includegraphics{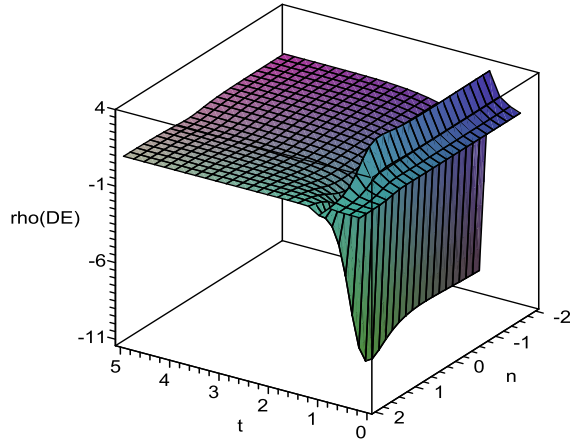} \vspace{7cm}
\end{center}
 \caption{\small {Variation of the effective dark energy versus
$n$ and the cosmic time.}}
\end{figure}

\subsection{The Expansion History}
To investigate expansion history of our model and comparing it with
other alternative theories, we study luminosity distance versus
redshift in this scenario. For a dark energy model with constant
equation of state parameter, the luminosity distance versus redshift
can be expressed as follows
\begin{equation}
d_{L}^{\omega}(z)=(1+z)\int_{0}^{z}\frac{H_{0}^{-1}dz}{\bigg[\Omega_{M}^{\omega}(1+z)^{3}+
(1-\Omega_{M}^{\omega})(1+z)^{3(1+\omega)}\bigg]^{1/2}}\,\,,
\end{equation}
and for a LCDM model, $\omega=-1$.  Now, the evolution of the cosmic
expansion in our DGP-inspired $F(R,\phi)$ model is given by
\begin{equation}
\frac{H(z)}{H_{0}}=\frac{1}{2}\Bigg[-\frac{1}{r_{0}H_{0}}+
\sqrt{\Big(2+\frac{1}{r_{0}H_{0}}\Big)^{2}+4\Omega_{M}^{0}
\Big[(1+z)^{3}-1\Big]+4\Omega_{\phi}^{0}[(1+z)^{3(1+\omega)}-1]}\,\,\Bigg],
\end{equation}
where by definition $r_{0}=\ell_{DGP}F'(R,\phi)$. The luminosity
distance versus redshift in a LDGP model can be expressed as [22]
\begin{equation}
d_{L}^{LDGP}(z)=(1+z)\int_{0}^{z}\frac{dz}{H(z)},
\end{equation}
and in our DGP-inspired $F(R,\phi)$ scenario, this quantity denoted
as $d_{L}^{FDGP}(z)$ is given by
\begin{equation}
d_{L}^{FDGP}(z)=(1+z)\int_{0}^{z}\frac{dz}{H(z)},
\end{equation}
where $H(z)$ is given by equation (46). Figure $7$ shows a
comparison between expansion histories of LCDM, LDGP and our FDGP
scenario for
$F(R,\phi)=\frac{1}{2}(1-\xi\phi^{2})[R-(1-n)\zeta^{2}(R/\zeta^{2})^{n}]
$ with $\xi=1/6$ and $n=0.8$. Note that this value of $n$ lies in
the appropriate range required for realization of the phantom-like
effect obtained in the previous subsection and it is also suitable
for describing late-time acceleration ( see for instance the paper
by Sotiriou and Faraoni in Ref. [15]). A LCDM scenario has very good
agreement with recent observations. As we see here, the FDGP
scenario is closer to LCDM more than LDGP. In other words, FDGP has
better agreement with recent observation than LDGP. Therefore FDGP
provides a better framework for treating phantom-like cosmology
without introducing any phantom field. Since we have not introduced
any phantom matter on the brane ( $\phi$ is a quintessence field
which plays the role of standard matter on the brane), it seems that
the null energy condition should be respected in this setup.
However, as we will show in subsection $3.6$, this is valid only for
some specific values of the model parameters and only in some
subspaces of the model parameter space.

\begin{figure}[htp]
\begin{center}\includegraphics{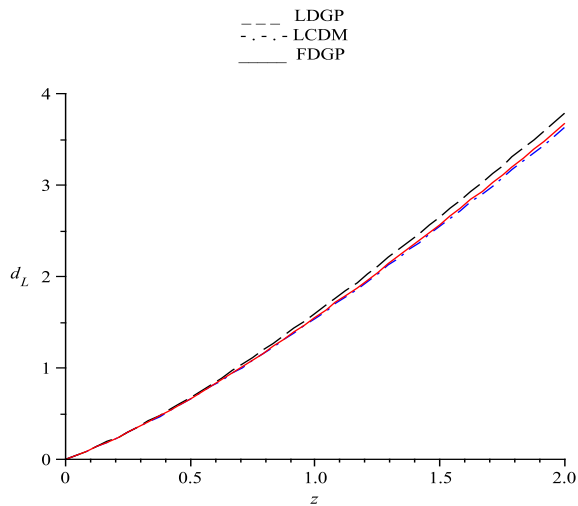} \vspace{6cm}
\end{center}
 \caption{\small { Luminosity distance versus redshift for three
 alternative scenarios. For a model universe with
 $F(R,\phi)=\frac{1}{2}(1-\xi\phi^{2})[R-(1-n)\zeta^{2}(R/\zeta^{2})^{n}]
$ with $\xi=1/6$ and $n=0.8$, there is better fit between FDGP and
LCDM. }}
\end{figure}

\subsection{Dynamics of the Equation of State Parameter}
To have more detailed discussion on the cosmological dynamics in
this model, we find from equation (40) the following relation ( with
$8\pi G=1$)
\begin{equation}
H=-\frac{1}{2r_{0}}\Bigg(1-\sqrt{1+4r_{0}^{2}\Big(\frac{\rho_{m}+\rho_{\phi}}{3}+\frac{\Lambda}{3}\Big)}\,\Bigg).
\end{equation}
Considering the energy conservation equation which is expressed here
as $\dot{\rho}_{tot}+3H(\rho_{tot}+p_{tot})=0$ where
$\rho_{tot}=\rho_{m}+\rho_{\phi}$ and $p_{tot}=p_{m}+p_{\phi}$, we
find
\begin{equation}
\dot{H}=-\frac{1}{2}\Bigg[\rho_{tot}(1+\omega_{tot})-
\frac{\dot{F}'(R,\phi)}{\ell_{DGP}[F'(R,\phi)]^{2}}\Bigg]
\Bigg(1-\frac{1}{\sqrt{1+4[F'(R,\phi)]^{2}l_{DGP}^2(\frac{\rho_{tot}}{3}+
\frac{\Lambda}{3})}}\Bigg).
\end{equation}
There is no superacceleration in this DGP-inspired $F(R,\phi)$
scenario if the following condition holds
\begin{equation}
\rho_{tot}\Big(1+\omega_{tot}\Big)>
\frac{\dot{F}'(R,\phi)}{\ell_{DGP}[F'(R,\phi)]^{2}}.
\end{equation}
To have a general relativistic interpretation of the expansion
history of this model, we rewrite the energy conservation equation
as follows
\begin{equation}
\dot{\rho}_{eff}+3H(1+\omega_{eff})\rho_{eff}=0
\end{equation}
and using equation (40) we have
\begin{equation}
\dot{\rho}_{eff}=\frac{-3\Big[F'(R,\phi)\Big]^{-1}\dot{H}}{\ell_{DGP}}+
\frac{3H\Big[F'(R,\phi)\Big]^{-2}\dot{F}'(R,\phi)}{\ell_{DGP}}.
\end{equation}
By comparison of equations (52) and (53), we find
\begin{equation}
1+\omega_{eff}=\frac{\Big[F'(R,\phi)\Big]^{-1}\dot{H}}{H
\ell_{DGP}\rho_{eff}}-
\frac{\Big[F'(R,\phi)\Big]^{-2}\dot{F}'(R,\phi)}{\ell_{DGP}\rho_{eff}}
\end{equation}
To realize the phantom phase in this DGP-inspired $F(R,\phi)$ model,
the condition $1+\omega_{eff}<0$ should be fulfilled. This leads us
to the following condition:
\begin{equation}
\frac{\dot{H}}{H}<\frac{\dot{F}'(R,\phi)}{F'(R,\phi)}.
\end{equation}
It is obvious that this model has the potential to describe the
crossing of the phantom divide line.

\subsection{The Null Energy Condition}

It is important to check the validity of the null energy condition
in this setup. In fact, the main feature of this setup is the
realization of the phantom-like behavior without introducing any
phantom matter on the brane. The null energy condition is respected
if the condition $\rho+p>0$ is valid. In our case, this condition is
given by $\rho_{tot}+p_{tot}>0$ where $\rho_{tot}$ and $p_{tot}$ are
defined in the subsection $3.2$. Figure $8$ shows the variation of
$y\equiv (\rho+p)_{tot}$ versus $n$ for some specific values of
redshift. As this figure shows, there are appropriate subspaces of
the model parameter space that the null energy condition is
respected in this setup. This is enough to say that this
DGP-inspired $F(R,\phi)$ model realizes the phantom-like behavior
without violating the null energy condition, at least in some
subspaces of the model parameter space. For instance, at $z=0.25$ (
which is corresponding to the epoch of the phantom-divide line
crossing), the null energy condition is respected if $n\leq 1.8$.
Albeit, those values of $n$ are adequate that are supported
observationally( by, for instance, solar system tests). It should
however be noticed that this range seems more restrained at higher
redshifts. The reason for violation of the null energy condition in
some subspaces of the model parameter space lies in the fact that a
modified theory of gravity of the form $f(R)$ is equivalent to a
theory of standard gravity plus a scalar field. With $f(R)$ gravity,
we have shown that one can mimic a phantom-like behavior without
introduction of a phantom field, but when the scenario is written in
the Einstein frame, the resulting scalar will violate the null
energy condition. So, it is natural to accept that in our model
there are some subspaces of the model parameter space that the null
energy condition can be violated. The main achievement is however
the existence of other subspaces that respect the null energy
condition.

\begin{figure}[htp]
\begin{center}\includegraphics{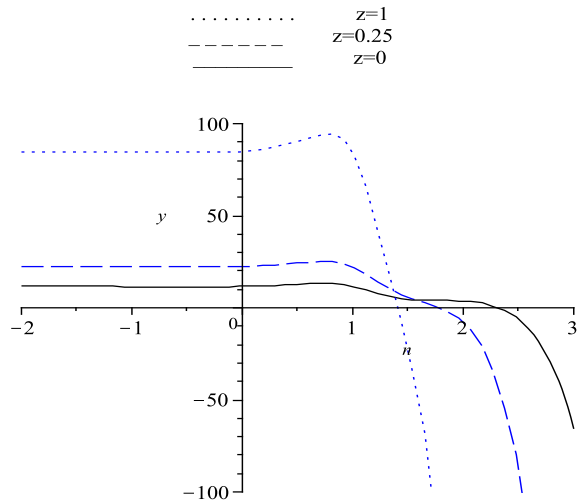} \vspace{6cm}
\end{center}
 \caption{\small { $y\equiv
(\rho+p)_{tot}$ versus $n$ for some specific values of redshift.
There are some subspaces of the model parameter space that null
energy condition is fulfilled for a model universe with
 $F(R,\phi)=\frac{1}{2}(1-\xi\phi^{2})[R-(1-n)\zeta^{2}(R/\zeta^{2})^{n}]
$ with $\xi=1/6$. }}
\end{figure}
\newpage
\section{Summary}
Based on the Lue-Starkman conjecture on the dynamical screening of
the brane cosmological constant in DGP scenario, in this paper we
have extended this proposal to a general DGP-inspired $F(R,\phi)$
Model. Firstly, we have studied phantom-like behavior in the normal
branch of an extension of DGP model where a quintessence field is
coupled non-minimally to the induced gravity on the brane. The
reason for incorporation of this canonical scalar field lies in the
fact that without scalar field it is impossible to realize phantom
divide line crossing in DGP setup. We have shown that the effective
dark energy density decreases by increasing the values of the
conformal coupling $\xi$ in a constant cosmic time slice. However,
for a constant $\xi$, we have phantom-like behavior ( increasing of
the effective dark energy density with cosmic time) in the normal
branch of the scenario without introducing any phantom field. Then
we have extended our study to a general DGP-inspired $F(R,\phi)$
scenario where we incorporate possible modification of the induced
gravity on the brane. In this case we obtained some new and
interesting results which we summarize as follows: by adopting the
ansatz
$F(R,\phi)=\frac{1}{2}(1-\xi\phi^{2})[R-(1-n)\zeta^{2}(R/\zeta^{2})^{n}]\,,
$ we have shown that phantom-like behavior can be realized in the
normal branch of the scenario if  $n\geq 0.73$ and $n\leq -0.60$. In
other words, for $-0.6\leq n \leq 0.73$ the effective dark energy in
this DGP-inspired $F(R,\phi)$ model has no phantom-like behavior.
Investigation of the expansion history of this model shows that this
DGP-inspired $F(R,\phi)$ scenario has the best fit with the recent
observational data. In fact this model is very close to a LCDM
scenario. Finally we found conditions for transition to phantom
phase of this model which has the potential to realize phantom
divide line crossing. For the case of a quintessence scalar field
non-minimally coupled to the induced gravity on the brane, the null
energy condition is fulfilled since there is no phantom matter on
the brane and the phantom dynamics is essentially gravitational
which saves the null energy condition. Also that the brane tension
does not violate the null energy condition too. For a general
DGP-inspired $F(R,\phi)$ scenario, the null energy condition is
respected only in some subspaces of the model parameter space
depending on the choice of the model of modified gravity.\\

{\bf Acknowledgement}\\
We would like to thank a referee for his/her important contribution
in this work.

\end{document}